\documentclass[amssymb,amsmath,superscriptaddress,footinbib]{revtex4}
\usepackage{amsmath,graphicx,subfigure,subeqnarray,fancyhdr,amsmath,multirow,color, mathrsfs,relsize}
\usepackage{natbib}
\begin{document}

\newcommand{\blue}[1]{{\color{black} #1}}

% Define new colors
\definecolor{pink}{rgb}{1.0, 0.11, 0.81}
\definecolor{cyan}{rgb}{0.0, 1.0, 1.0}
\definecolor{green}{rgb}{0.0, 0.5, 0.0}
\definecolor{purple}{rgb}{0.59, 0.44, 0.84}

\newcommand{\pd}[2]{\frac{\partial #1}{\partial #2}}
\def\abs{\operatorname{abs}}
\def\argmax{\operatornamewithlimits{arg\,max}}
\def\argmin{\operatornamewithlimits{arg\,min}}
\def\diag{\operatorname{Diag}}
\newcommand{\eqRef}[1]{(\ref{#1})}
\newcommand{\Rext}{R_\text{ext}}
\newcommand{\bext}{\beta_\text{ext}}
\newcommand{\erm}{\mathrm{e}}
\newcommand{\irm}{\mathrm{i}}
\newcommand{\drm}{\mathrm{d}}
\newcommand{\jrm}{\mathrm{j}}
\newcommand{\dphi}{\Delta\phi}
% symbols for Cauchy principal value integrals
\def\Xint#1{\mathchoice
   {\XXint\displaystyle\textstyle{#1}}%
   {\XXint\textstyle\scriptstyle{#1}}%
   {\XXint\scriptstyle\scriptscriptstyle{#1}}%
   {\XXint\scriptscriptstyle\scriptscriptstyle{#1}}%
   \!\int}
\def\XXint#1#2#3{{\setbox0=\hbox{$#1{#2#3}{\int}$}
     \vcenter{\hbox{$#2#3$}}\kern-.5\wd0}}
\def\ddashint{\Xint=}
\def\dashint{\Xint-}

\title{Electro-hydrodynamic synchronization of piezoelectric flags}


\author{Yifan Xia}
\affiliation{LadHyX--D\'epartement de M\'ecanique, CNRS -- \'Ecole Polytechnique, Route de Saclay, 91128 Palaiseau, France}
\author{Olivier Doar\'e}
\affiliation{ENSTA, Paristech, Unit\'e de M\'ecanique (UME), Chemin de la Huni\`ere, 91761 Palaiseau, France}
\author{S\'ebastien Michelin}
\email{sebastien.michelin@ladhyx.polytechnique.fr}
\affiliation{LadHyX--D\'epartement de M\'ecanique, CNRS -- \'Ecole Polytechnique, Route de Saclay, 91128 Palaiseau, France}

\date{\today}
\begin{abstract}
Hydrodynamic coupling of flexible flags in axial flows may profoundly influence their flapping dynamics, in particular driving their synchronization. This work investigates the effect of such coupling on the harvesting efficiency of coupled piezoelectric flags, that convert their periodic deformation into an electrical current. Considering two flags connected to a single output circuit, we investigate using numerical simulations the relative importance of hydrodynamic coupling to electrodynamic coupling of the flags through the output circuit due to the inverse piezoelectric effect. It is shown that electrodynamic coupling is dominant beyond a critical distance, and induces a synchronization of the flags' motion resulting in enhanced energy harvesting performance. We further show that this electrodynamic coupling can be strengthened using resonant harvesting circuits.

\end{abstract}

\maketitle

%\linenumbers

\section{Introduction}
Piezoelectric materials draw from their internal structure their fundamental ability to generate  a net charge displacement when they are deformed and to respond mechanically to an electric forcing. This two-way electro-mechanical coupling may be exploited to convert the mechanical energy of a vibrating structure into electrical form, and has become increasingly popular to design energy harvesters based on ambient vibrations~\citep{allen:2001,sodano:2004, anton:2007, erturkbook:2011, calio:2014piezoelectric}.

Such systems critically depend on existing or forced vibrations of the structure. The concept can nevertheless be extended to harvest energy from a steady flow by exploiting fluid-solid instabilities to generate self-sustained motions of a solid body~\citep{xiao:2014}. The spontaneous flapping of thin deformable plates in axial flow beyond a critical flow velocity is another example, commonly referred to as the flag instability (see the recent review in Ref.~\citep{shelley:2011}). This instability has been the focus of intense recent investigations to understand the impact of energy extraction on the flapping dynamics~\citep{singh2012effect} and assess the energy harvesting performance when the flag's deformation is converted into electric energy using piezoelectric and other electroactive materials covering the flag's surface~\citep{dunnmon:2011, giacomello:2011, akcabay:2012, michelin:2013,pineirua2015influence}.

Much of the work on piezoelectric flags has so far focused on single flags connected with simple circuits, such as pure resistors, to understand the effect of the fluid-solid-electric coupling on the system's stability and the energy transfers between the fluid, solid and electrical components~\citep{dunnmon:2011,doare:2011, akcabay:2012}. Ref.~\cite{michelin:2013} further showed that the flapping amplitude and frequency of a piezoelectric flag can be significantly modified by the extraction of energy and coupling to the dynamical properties of the output circuit even for a purely resistive output. Using a resonant circuit, Ref.~\citet{xia:2015} reported a frequency lock-in phenomenon that considerably increases the flag's flapping amplitude and efficiency: during lock-in, the flapping frequency of the flag is dictated by the circuit to match its resonance frequency therefore enabling large voltage and energy transfers to the output load. 

Because of its material properties, a single piezoelectric harvester is still characterized by its small power output \citep{sodano:2004, anton:2007}. One potential solution to this problem is to combine multiple devices in order to produce the required power. For flapping flags, however, the placement in close proximity of multiple flapping structures significantly modifies their flapping dynamics, and hydrodynamic synchronization as well as modification of the flapping amplitude and frequency have been reported in multiple recent studies~\citep{schouveiler:2009, alben:2009c, michelin:2009, michelin2010,mougel2016}. 

Depending on their relative positioning, two side-by-side flags may flap in-phase (with identical vertical displacements), or out-of-phase (with opposite vertical displacement)~\citep{zhang:2000, zhu2003interaction}. More complex synchronization was also identified in numerical simulations for both side-by-side  and tandem flags~\citep{alben:2009c,tian2011}. The synchronization of the flags can modify their individual dynamical properties and their individual performance as energy harvesters. Furthermore, if the flags are to be connected electrically to a single device, \blue{for example to increase the available power}, the relative phase and amplitude of the generated signals (directly related to their mechanical dynamics) will be critical: the electric interaction might be constructive (resp. destructive) if both signals are in-phase (resp. out-of-phase). Hydrodynamic coupling therefore influences the efficiency of flags that are electrically-isolated~\citep{song2014energy}. Finally, the inverse piezoelectric effect introduces a feedback forcing on the flags' dynamics by the electrical circuit: when multiple flags are connected to the same output loop, this introduces an additional electrodynamic coupling that competes with hydrodynamic effects in setting the relative phase and dynamical properties of the flapping motion.

The motivation for the present work is therefore to investigate the role and relative weight of these different coupling mechanisms and the resulting harvesting efficiency of coupled piezoelectric flags. To this end, we focus on the system consisting of two flags placed side-by-side in an axial flow and connected to a single output circuit, \blue{a simple fluid-solid-electric system which provides physical insight on the hydro- and electro-dynamic couplings and performance of a two-flag assembly.} In Section~\ref{sec:model}, the models used to describe the fluid-solid-electric dynamics are presented. Section~\ref{sec:results} analyses the relative importance of hydrodynamic and electrodynamic coupling in synchronizing the flags' motion. The role of the output circuit is then discussed in Section~\ref{sec:circuit}. Finally, conclusions and perspectives are presented in Section~\ref{sec:conclusions}.

\section{Fluid-solid-electric model}\label{sec:model}
\begin{figure}[t]
\centering
\includegraphics[width=.6\textwidth]{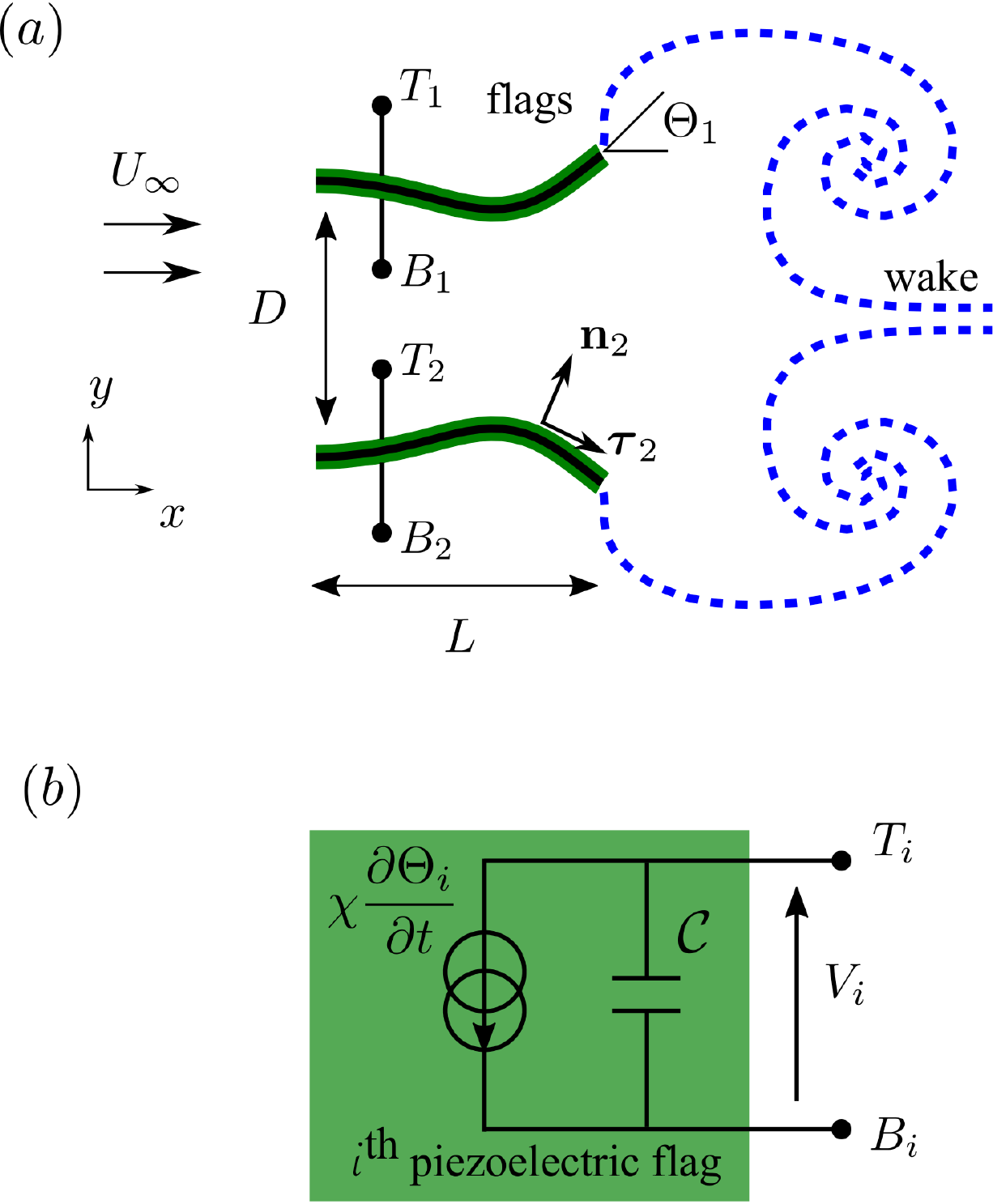}
\caption{\label{fig:notations}$(a)$ A schematic representation of two piezoelectric flags placed side-by-side in an axial flow, and $(b)$ the electric equivalence of the $i$\textsuperscript{th} piezoelectric flag with its two electrodes. The flags' electrodes are connected in a same circuit.}
\end{figure}

We consider here two piezoelectric flags placed side by side in an axial flow. The flapping motion of these structures are coupled both hydrodynamically (each flag modifies the flow field experienced by the other one) and electrodynamically (both flags are connected to the same output circuit). A schematic representation of the coupled system is shown in Fig.~\ref{fig:notations}.

\subsection{Piezoelectric flags modeling}
Both piezoelectric flags are clamped side by side from their leading edge and placed in an axial fluid flow of density $\rho_f$ and velocity $U_\infty$ at a distance $D$ from each other. Both flags are infinitely thin and have a span-wise dimension $H$ much larger than their streamwise length $L$, so that the flags' and fluid's motions are purely two-dimensional. Each flag is entirely covered by a single pair of piezoelectric electrodes shunted through the flag with reverse polarity. The piezoelectric material is characterized by an electromechanical coefficient $\chi=e_{31}h/2$, with $h$ the thickness of the piezoelectric flag and $e_{31}$ the reduced piezoelectric coupling coefficient~\citep{erturk:2009} and intrinsic capacitance $\mathcal{C}$. The resulting three-layer plate has bending rigidity $B$ per unit length and mass per unit area $\mu$. In the following, the problem is non-dimensionalized using $L$, $L/U_\infty$, $\mu L^2$ and $U_\infty\sqrt{\mu L/\mathcal{C}}$ respectively as characteristic length, time, mass and voltage scales. Fluid forces are naturally scaled by $\rho U_\infty^2$.

\blue{The deformation of flag $i$ ($i=1,2$) induces a charge displacement $Q_i$ within the corresponding piezoelectric pair, driven by the change in the patches' length. For a thin patch positioned on an inextensible flag, it is determined by the relative orientations of the flag centerline at both ends of the patch~\citep{doare:2011}. Here, the entire flag is covered by a single pair and the leading edge is clamped: noting $\Theta_i(s)$ the flag's orientation with respect to the flow direction, the charge displacement is therefore completely determined by $\Theta_i(s=L)$. For the $i$-th flag ($i=1,2$), the charge displacement $Q_i$ is then given by:} \citep{ducarne2012placement}
\begin{equation}
Q_i=\frac{\alpha}{U^*}\Theta_i(s=L)+V_i,\label{eq:dir_effect_ad}
\end{equation}
\blue{where $V_i$ is the voltage across the piezoelectric pair of flag $i$}, and $\alpha$ and $U^*$ are the piezoelectric coupling coefficient and relative velocity of the fluid flow and of structural bending waves, respectively defined as:
\begin{equation}
\label{eq:def_alpha}
\alpha=\chi\sqrt{\frac{L}{B\mathcal{C}}}, \quad U^*=U_\infty L\sqrt{\frac{\mu}{B}}.
\end{equation}
The inverse piezoelectric effect converts the electric field within the piezoelectric material into mechanical stress. For \blue{a pair of thin patches covering the entire flag}, this leads to a piezoelectric torque \citep{preumont:2011} applied at the trailing edge \blue{flag $i$}:
\begin{equation}
\mathcal{M}_\text{piezo}=-\frac{\alpha}{U^*} V_i.\label{eq:inv_effect_ad}
\end{equation}

The dynamics of each flag is described using the inextensible Euler-Bernoulli model, which is written in dimensionless form as:
\begin{align}
\label{eq:EB_dimensionless}
\frac{\partial^2\mathbf{x}_i}{\partial t^2}&=\frac{\partial}{\partial s}\left(T_i\boldsymbol{\tau}_i-\frac{1}{U^{*2}}\frac{\partial^2\Theta_i}{\partial s^2}\mathbf{n}_i\right)
-M^*\Delta p_i\mathbf{n}_i,\\
\label{eq:inextens}
\frac{\partial\mathbf{x}_i}{\partial s}&=\boldsymbol{\tau}_i,
\end{align}
where \blue{$T_i(s,t)$ is the tension distribution within flag $i$} and \blue{$\Delta p_i=p_+-p_-$ is the difference of fluid pressure applied on the upper surface ($p_+$) and lower surface ($p_-$) of the flag. Note that, for an inextensible flag, $T_i(s,t)$ is not determined \emph{a priori} but instead plays the role of a Lagrange multiplier to guarantee the inextensibility condition.} The flags are clamped at their leading edge and free at their trailing edge:
\begin{align}
&\text{at}\quad s=0:~\blue{\Theta_i=0,\quad \mathbf{x}_1=0,\quad \mathbf{x}_2=d\,\mathbf{e}_y}\label{eq:BC_adim_1},\\
&\text{at}\quad s=1:~T_i=\frac{1}{U^{*2}}\frac{\partial \Theta_i}{\partial s}-\frac{\alpha}{U^*} V_i=\frac{1}{U^{*2}}\frac{\partial^2\Theta_i}{\partial s^2}=0.\label{eq:BC_adim_2}
\end{align}
In particular, Eq.~\eqref{eq:BC_adim_2} indicates that the \blue{tension, internal torque and shear force must vanish at the free trailing edge, including both the elastic components and the additional piezoelectric torque applied locally on $s=1$}.  The fluid-solid problem depends on two additional non-dimensional parameters:
\begin{equation}
\label{eq:def_MUd}
M^*=\frac{\rho_fL}{\mu},\quad d=\frac{D}{L},
\end{equation}
that are namely the fluid-solid mass ratio and the relative distance between the flags. %Note also that all vectors involved in the models are represented using complex numbers.

\subsection{Vortex sheet model}
Solving for the fluid-solid dynamics requires a model for the fluid force $\mathbf{f}^\textrm{fluid}_i$ in Eq.~\eqref{eq:EB_dimensionless}. We assume here that the fluid's viscosity is negligible except within thin boundary layers along the flag that separate at the trailing edge and roll up into vortical structures. To describe this essentially-inviscid flow, we adopt a potential flow model, describing each flag as a bound vortex sheet (i.e. a discontinuity in tangential velocity) and the vortex wake as a free vortex sheet continuously shed from the flag's trailing edge, thereby applying the vortex sheet model introduced and presented in details by Ref.~\citet{alben:2009a}. 

The flow is inviscid, incompressible and irrotational except for two vortex sheets $C_1$ and $C_2$.  $C_i$ includes both the bound vortex sheet attached to the flag ($0\leq s\leq 1$) and the free vortex sheet shed at the flag's edge ($s>1$). Using Biot-Savart's law, the fluid's velocity field at $\mathbf{x}$ is then obtained as:
\begin{equation}
\label{eq:2flags_velocity}
\mathbf{u}(\mathbf{x},t)=\mathbf{e}_x+\sum_{i=1,2}\frac{1}{2\pi}\int_{C_i}\frac{\mathbf{\gamma}_i(s',t)\mathbf{k}\times[\mathbf{x}-\mathbf{x}_i(s',t)]}{|\mathbf{x}-\mathbf{x}_i(s',t)|^2}\drm s',
\end{equation}
where $\gamma_i(s,t)$ is the local strength of the vortex sheet $C_i$. When $\mathbf{x}$ is on $C_i$, the previous expression remains valid for the mean flow velocity $\tilde{\mathbf{u}}_i$ (i.e. the average of the flow velocity on either side of the sheet) provided that the integral on $C_i$ is understood as a Cauchy Principal Value Integral. The free vortex sheet is simply advected by the local flow field:
\begin{equation}\label{eq:advection}
\pd{\mathbf{x}_i}{t}(s,t)=\tilde{\mathbf{u}}_i(\mathbf{x}_i(s,t),t),\qquad s>1,
\end{equation}
 and \blue{the circulation, {\it i.e.} the arc-length integral of $\gamma_i$ along $C_i$,} acts as a passive tracer for $s\geq 1$. Enforcing the \blue{continuity of the} normal flow velocity on each flag provides coupled singular integral equations for the intensities $\gamma_i(s,t)$ of the bound vortex sheets:
\begin{equation}
\label{eq:impermeability}
\mathbf{n}_i\cdot\pd{\mathbf{x}_i}{t}(s,t)=\mathbf{n}_i\cdot\mathbf{u}(\mathbf{x}_i(s,t),t),\quad 0\leq s\leq 1.
\end{equation}
The regularity of the flow field near the trailing edge of each flag and conservation of total circulation around $C_i$ provides a unique solution for the integral equation problem.

In this inviscid limit, the fluid force $\mathbf{f}^\textrm{fluid}_i$ comes purely from the difference of pressure on either side of the flag and is obtained from Bernoulli's theorem \citep{michelin:2008, alben:2009a}:
\begin{equation}
\label{eq:pressure_Bernoulli}
\Delta p_i(s,t)=\left[\int_{0}^s\pd{\gamma_i}{t}+\gamma_i\boldsymbol\tau_i\cdot\left(\tilde{\mathbf{u}}_i-\pd{\mathbf{x}_i}{t}\right)\right].
\end{equation}

\subsection{Electrical circuits}
\begin{figure}
\begin{center}
\includegraphics[width=.7\textwidth]{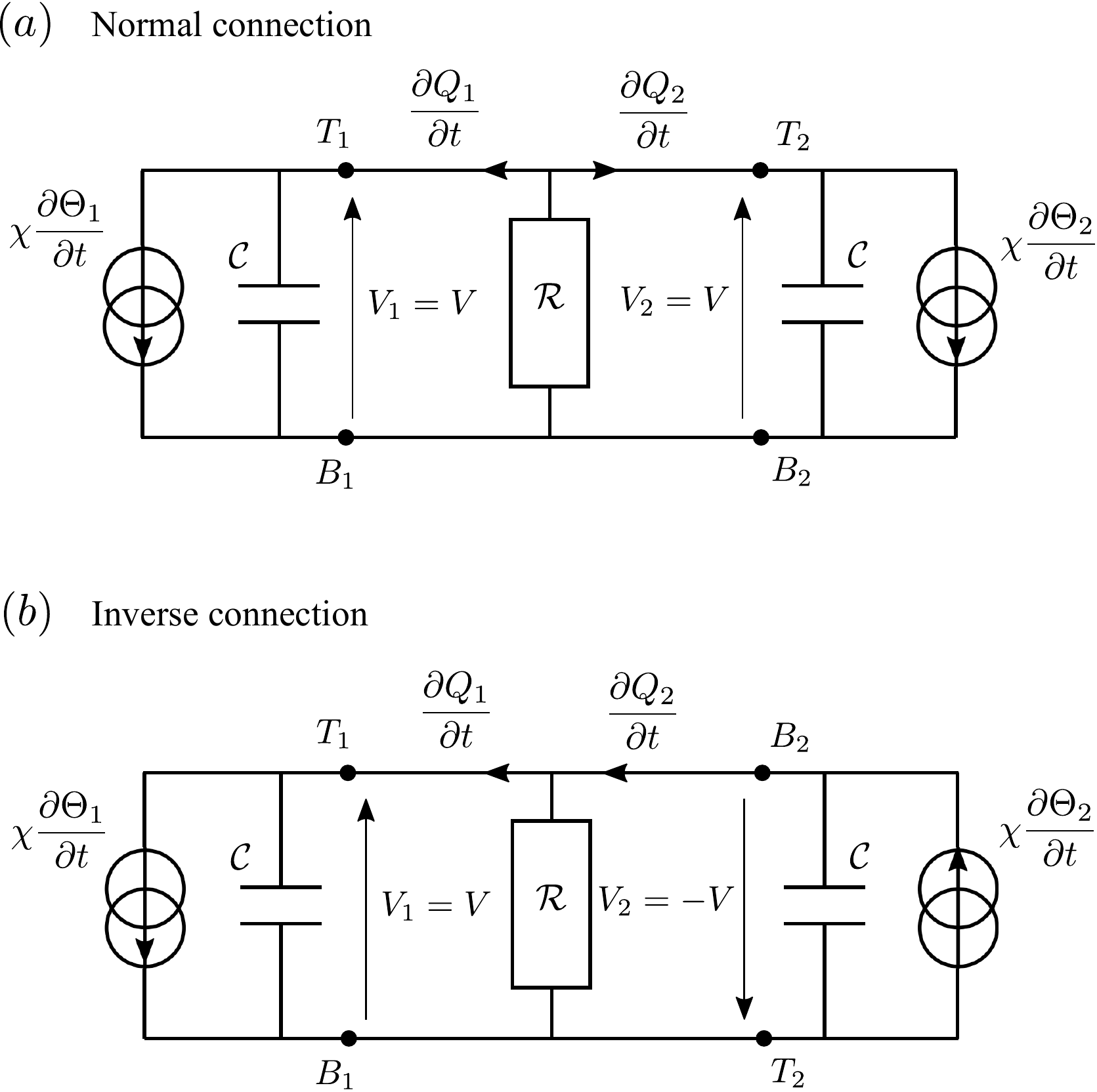}
\caption{\label{fig:circuit_R}Resistive circuit in $(a)$ normal connection and $(b)$ inverse connection}
\end{center}
\end{figure}

Two electrodes stretch out from each flag and are respectively denoted as $T$ and $B$ (see Fig.~\ref{fig:notations}$b$), with the index 1 and 2 distinguishing flags 1 and 2. We may consider two types of connections: (i) the {\it normal connection} (NC) by joining $T_1$ to $T_2$, and $B_1$ to $B_2$ (Fig.~\ref{fig:circuit_R}$a$), or (ii) the {\it inverse connection} (IC) by joining $T_1$ to $B_2$, and $B_1$ to $T_2$ (Fig.~\ref{fig:circuit_R}$b$).

The flags are connected in parallel to the same output circuit, thus sharing the same voltage ($V_1=V_2=V$) when the {\it normal connection} is used, and opposite voltage ($V_1=-V_2=V$) when the {\it inverse connection} is used. The dynamics of the output circuit can be written generally as
\begin{equation}
\label{eq:one_circuit}
V+\mathcal{Z}(\frac{\partial Q_1}{\partial t}\pm \frac{\partial Q_2}{\partial t})=0,
\end{equation}
where $\mathcal{Z}$ is formally the impedance of the output circuit, and the $+$ (resp.~$-$) sign in the previous equation corresponds to the normal (resp. inverse) connection. For a linear output circuit, $\mathcal{Z}$ is a linear integro-differential operator in time. For example, the case of a pure resistor $\mathcal{R}$ \blue{simply corresponds to }$\mathcal{Z}=\beta$, with $\beta$ the non-dimensional resistance of the output circuit, 
\begin{equation}
\beta=\frac{\mathcal{R}\mathcal{C}U_\infty}{L}.
\end{equation}
Unless specified otherwise, the {\it normal connection} is used here with a resistive circuit (Fig.~\ref{fig:circuit_R}$a$). Equation~\eqref{eq:one_circuit} is therefore rewritten as:
\begin{equation}
\label{eq:normal_resistive}
V+\beta(\frac{\partial Q_1}{\partial t}+ \frac{\partial Q_2}{\partial t})=0.
\end{equation}

\subsection{Energy harvesting efficiency}
The harvesting component of the output circuit (i.e. the useful load) is modeled as a resistor $\mathcal{R}$ \citep{dunnmon:2011, doare:2011}. In analogy with the single flag problem and the classical definition of efficiency for wind-turbines, the energy-harvesting efficiency $\eta$ of the two-flag assembly is evaluated as the ratio between the average power dissipated in the circuit's resistance $\langle\mathcal{P}\rangle$ to the fluid kinetic energy flux ($\rho U_\infty^3/2$) through the surface effectively swept by each flag's trailing edge:
\begin{equation}
\label{efficiency}
\eta=\frac{\langle\mathcal{P}\rangle}{\frac{1}{2}\left(\mathcal{A}_1+\mathcal{A}_2\right)},
\end{equation}
where $\mathcal{A}_i$ the dimensionless peak-to-peak flapping amplitude of flag $i$ at the trailing edge. The instantaneous rate of dissipation within the resistance is given, in dimensionless form, by $\mathcal{P}=V^2/\beta$.

\subsection{Numerical solution}
Equations~\eqref{eq:dir_effect_ad}, \eqref{eq:EB_dimensionless}--\eqref{eq:BC_adim_2}, \eqref{eq:2flags_velocity}--\eqref{eq:pressure_Bernoulli} and \eqref{eq:one_circuit} form a nonlinear system of partial differential equations for the fully-coupled fluid-solid-electric dynamics. This system is marched in time using a second-order semi-implicit time-stepping scheme. At each time step, the problems for the state variables (flag's curvature, voltage and circulation) is recast as a set of nonlinear equations which is solved iteratively \citep{broyden:1965}. A Chebyshev collocation method is used to compute derivations and integration in space~\citep{alben:2009a}.

The singularity of the first kernel in Eq.~\eqref{eq:2flags_velocity} is regularized as~$(\mathbf{k}\times\mathbf{r})/(|\mathbf{r}|^2+\delta^2)$ on the free vortex sheet to ensure proper numerical integration~\citep{krasny1986desingularization}, using $\delta=0.2$~\citep{alben:2008c}.

\section{Electrodynamic and hydrodynamic synchronizations}\label{sec:results}
In this section, we focus on the normal connection of the two piezoelectric flags with the simplest harvesting circuit, namely a single resistor $\mathcal{R}$ (Figure~\ref{fig:circuit_R}). \blue{Energy can only be harvested when the flags are unstable  and undergo spontaneous flapping; in the following, we therefore focus on this postcritical regime.}

Equations~\eqref{eq:dir_effect_ad} and \eqref{eq:one_circuit} now become:
\begin{equation}
\label{eq:circuit_R_NC}
2\pd{V}{t}+\frac{V}{\beta}+\frac{\alpha}{U^*}\left( \pd{\Theta_1}{t}+\pd{\Theta_2}{t} \right)=0.
\end{equation}
In the normal connection, the voltage between the upper and lower electrodes is identical for each flag, and so are the piezoelectric torques applied at each flag's trailing edge: electrodynamic coupling of the piezoelectric flags is therefore expected to favor an in-phase flapping pattern.

\subsection{Hydrodynamic synchronization with no piezoelectric coupling}
\begin{figure}[t]
\begin{center}
\includegraphics[width=\textwidth]{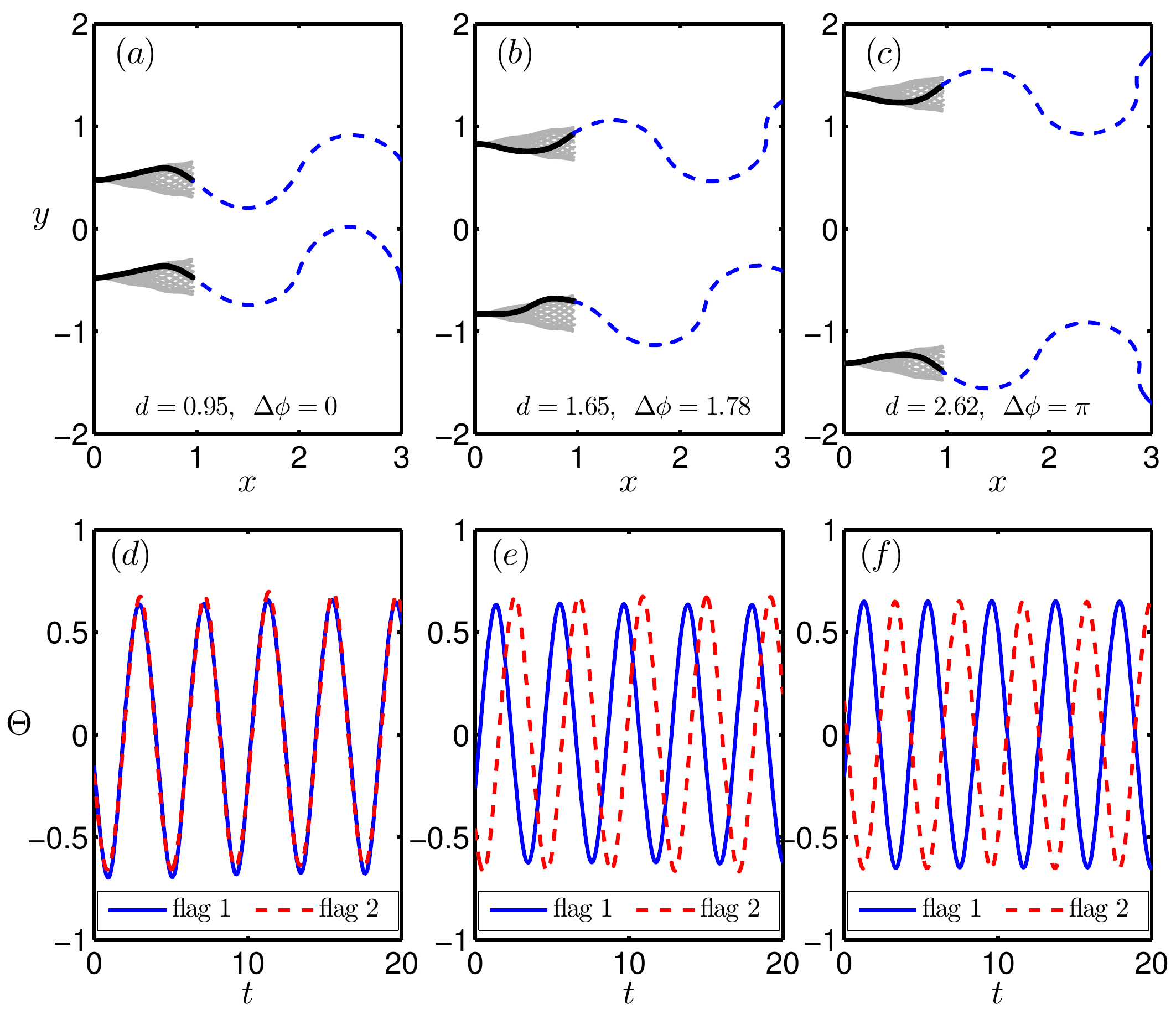}
\caption{\label{fig:M5U10_alpha0_3dis_3dphase}$(a,b,c)$ Flapping motion and instantaneous position of the flags and vortex sheets and $(d,e,f)$ evolution of trailing edge deflection $\Theta_i$ in time for $d=0.95$ (a,d), $d=1.65$ (b,e) and $d=2.62$ (c,f) for $\alpha=0$, $M^*=5$, $U^*=10$.}
\end{center}
\end{figure}

The flapping pattern without piezoelectric coupling ($\alpha=0$) is first investigated. Ref.~\cite{alben:2009c} reported a continuous evolution of the phase shift $\dphi$ with varying $d$. This observation is confirmed in our work using the same model in the absence of piezoelectric coupling (Fig.~\ref{fig:M5U10_alpha0_3dis_3dphase}): increasing $d$, the flapping pattern evolves continuously from in-phase flapping ($d=0.95$, Fig.~\ref{fig:M5U10_alpha0_3dis_3dphase}$a,d$) to out-of-phase flapping (Fig.~\ref{fig:M5U10_alpha0_3dis_3dphase}$c,f$). Rather than a sharp transition, the evolution from in-phase to out-of-phase is continuous (see intermediate phase for $d=1.65$, Fig.~\ref{fig:M5U10_alpha0_3dis_3dphase}$b,e$). Increasing $d$ further, the phase shift maintains this smooth and monotonic evolution \citep{alben:2009c}.

\subsection{Effect of piezoelectric coupling}
We now focus on the modification introduced to this hydrodynamic synchronization when the piezoelectric coupling is activated ($\alpha\neq 0$). All other parameters being held fixed, for a single flag, an optimal energy harvesting is obtained when the output circuit is perfectly tuned, i.e. when the time scales of the circuit and of the flapping are identical  \citep{michelin:2013}. For simplicity, the value of $\beta$ is chosen here for a perfect tuning for a single flag. 

\begin{figure}[t]
\begin{center}
\includegraphics[width=\textwidth]{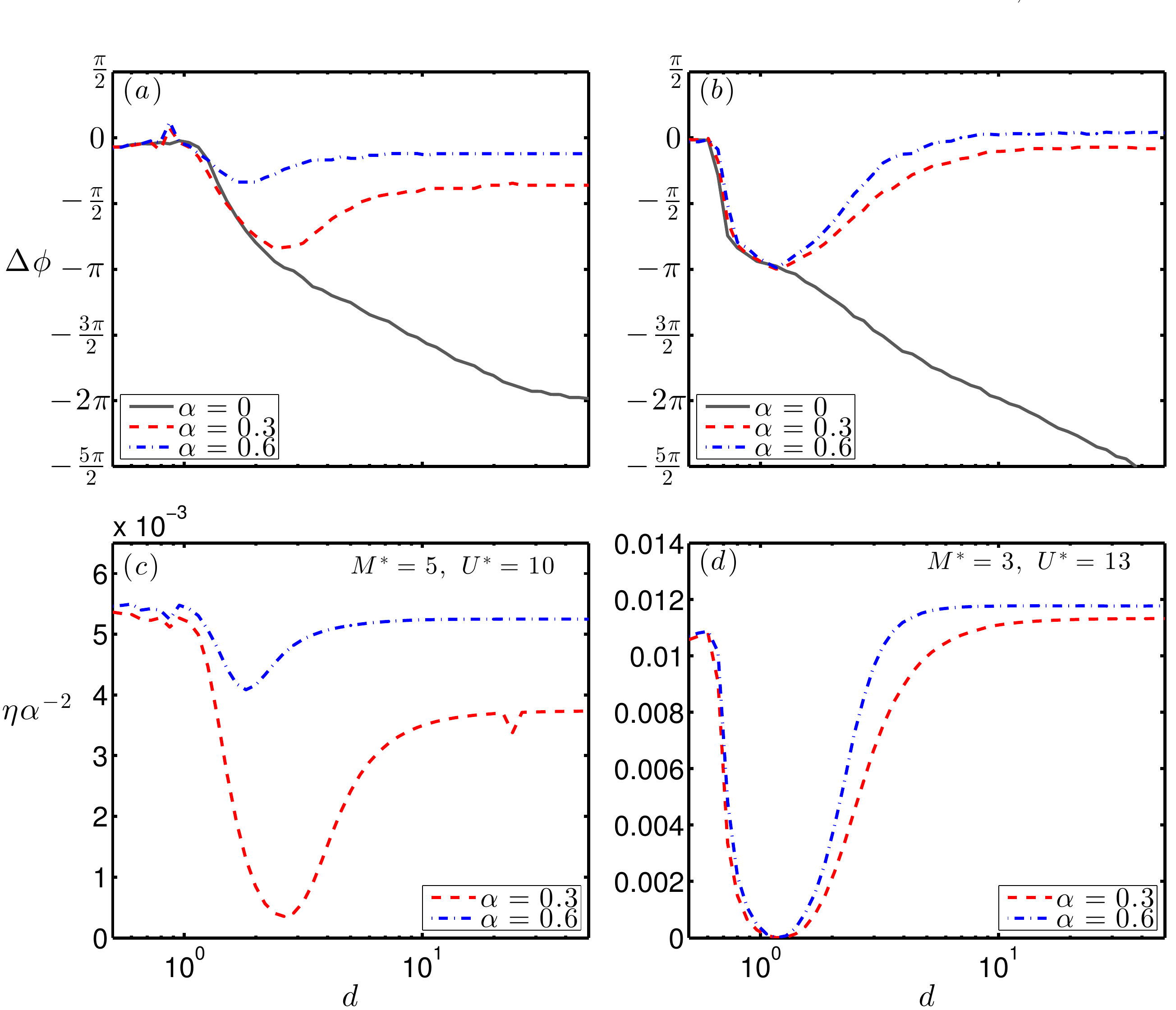}
\caption{\label{fig:para_M5_M3_phase_eff_dis}Evolution with $d$ of $(a,b)$ the relative phase of the flags $\dphi$, and $(c,d)$ the rescaled harvesting efficiency $\eta\alpha^{-2}$, for $M^*=5$, $U^*=10$, $\beta=0.17$ ($a,c$) or $M^*=3$, $U^*=13$, $\beta=0.15$ ($b,d$).}
\end{center}
\end{figure}
For small values of $d$, the evolution of $\dphi$ remains unchanged when $\alpha\neq 0$ (Fig.~\ref{fig:para_M5_M3_phase_eff_dis}$a,b$), demonstrating the dominance of hydrodynamic coupling over the piezoelectric coupling in synchronizing the flags. The energy harvesting efficiency is nevertheless strongly impacted by the phase shift: in-phase flapping results in ${Q}_1\approx{Q}_2$ and the forcing of the flags on the circuit are additive resulting in high efficiency. The efficiency decreases as $\dphi$ evolves towards $-\pi$ (out-of-phase flapping). In that case, the forcing of the two flags on the circuit cancels out, leading to a very small current in the output resistor.

For larger separation distances, the electrodynamic coupling modifies the synchronization, in favor of an in-phase dynamics ($\dphi\rightarrow 0$) as anticipated  (see for example, Fig.~\ref{fig:para_M5_M3_phase_eff_dis}$b$). As a result, the harvesting efficiency at large distance is more important and comparable to the one obtained for in-phase flapping at small distance (see Fig.~\ref{fig:para_M5_M3_phase_eff_dis}$d$). Those results also show a scaling of $\eta$ as $\alpha^2$ in the efficient regime as explained by Ref.~\citet{michelin:2013} for small electromechanical coupling. 

\subsection{Hydrodynamic vs. electrodynamic forcing}
\label{sec:scaling}
These observations are consistent with the weakening of hydrodynamic coupling with increasing $d$, while electrodynamic coupling is independent of $d$.  A characteristic distance $d_c$ therefore exists where both effects are of the same order and a transition from hydrodynamic to electrodynamic synchronization occurs.

We seek a scaling law for $d_c$ by comparing the relative magnitude of the piezoelectric and hydrodynamic torques generated on one of the flag by the second one. In dimensional form, the former, $\mathcal{M}_\text{piezo}$, scales as:
\begin{equation}
\label{eq:scale_piezo_torque}
\mathcal{M}_\text{piezo}\sim \chi [V]\sim \frac{\chi^2[\Theta]}{\mathcal{C}},
\end{equation}
where $[\Theta]=\mathcal{O}(1)$ is the typical trailing edge angle. The typical perturbation velocity introduced by one flag near the other is $[u]\sim U_\infty L/D$ (Biot-Savart's law), therefore the hydrodynamic torque created on one flag by the motion of the other, $\mathcal{M}_\text{fluid}\sim\rho_f[u]^2L^2$, scales as
\begin{equation}
\label{eq:scale_fluid_force}
\mathcal{M}_\text{fluid}\sim \rho_f L^2U_\infty^2\frac{L^2}{D^2}.
\end{equation}
Consequently, 
\begin{equation}
\label{eq:MpoverMf_1}
\frac{\mathcal{M}_\text{piezo}}{\mathcal{M}_\text{fluid}}\sim\frac{\chi^2[\Theta]}{\rho_fL^2U_\infty^2\mathcal{C}}\left(\frac{D}{L}\right)^2\sim\left(\frac{D}{D_c}\right)^2,\qquad \textrm{with   }D_c=\frac{LU_\infty}{\chi}\sqrt{\rho_f \mathcal{C}},
\end{equation}
or in dimensionless form
\begin{equation}\label{eq:critical_d_alpha}
d_c=\frac{U^*\sqrt{M^*}}{\alpha}.
\end{equation}
When $d\ll d_c$, the hydrodynamic coupling dominates and synchronization of the two flags follows closely the behavior obtained with no piezoelectric coupling. When $d\gg d_c$, electrodynamic coupling through the piezoelectric electrodes and circuit dominates and imposes in-phase flapping. 

Equation~\eqref{eq:critical_d_alpha} indicates that $d_c$ is proportional to $\alpha^{-1}$, consistent with the fact that a stronger electromechanical coupling coefficient results in an electrodynamic synchronization at smaller distance as piezoelectric effects are larger (Fig.~\ref{fig:para_M5_M3_phase_eff_dis}). 

\subsection{Synchronization effects: linear analysis}
\begin{figure}
\centering
\includegraphics[width=1\textwidth]{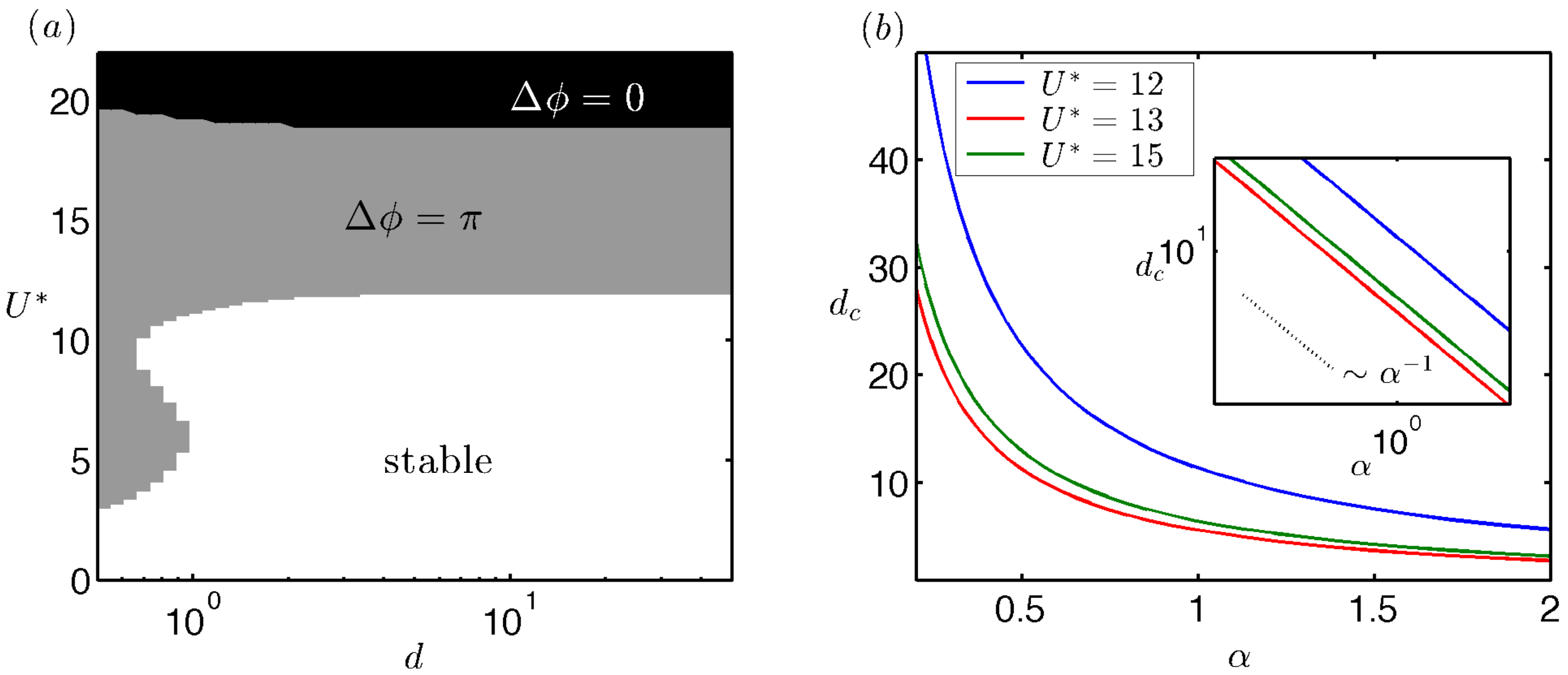}
\caption{\label{fig:carte_phase_M3_Uvar_Dvar_a0} $(a)$ Relative phase of the two flags $\dphi$ in the most unstable mode as a function of $d$ and $U^*$, for $M^*=3$, $\alpha=0$ and $\beta=0.1$. $(b)$ Evolution with $\alpha$ of the critical distance $d_c$ at which the synchronization of the most unstable mode switches from out-of-phase to in-phase, for $M^*=3$ and varying $U^*$. The same data is plotted in logarithmic scale as an inset.}
\end{figure}

To confirm the relation $d_c\sim\alpha^{-1}$, we investigate the synchronization properties of the linearly unstable modes of the system: the nonlinear equations for the fluid-solid-electric dynamics are linearized for small vertical displacements $y_i$ of the flags.

In this linear approximation, self-induction of the wake is negligible and so is the advection due to the plate's motion; therefore, the free vortex sheet is simply advected by the mean flow along the horizontal axis. The problem is therefore recast as an eigenvalue problem, which is nonlinear due to the velocity induced by the semi-infinite wake on the plate~\citep{alben:2008pof}. Because of the symmetries of the equations, eigenmodes of the problem are either in-phase ($y_1=y_2$, $Q_1=Q_2$) or out-of-phase ($y_1=-y_2$, $Q_1=-Q_2$)~\citep{michelin:2009}.

For $\alpha=0$ and $M^*=3$, Figure~\ref{fig:carte_phase_M3_Uvar_Dvar_a0}$a$ shows that the phase shift of the most unstable mode is $\dphi=\pi$ (out-of-phase) at the stability threshold. \blue{A reduction of the critical velocity beyond which spontaneous flapping develops can be observed at small distance $d$, which is consistent with previous work \citep{dessi2015aeroelastic}.} A switch to an in-phase dominant mode is then observed at higher velocity ($U^*>20$). This switch from out-of-phase to in-phase is observed for $\alpha\neq 0$ but for shorter distances as the coupling between the electric and solid system becomes stronger: the piezoelectric coupling increasingly favors the in-phase synchronization of the flags. Figure~\ref{fig:carte_phase_M3_Uvar_Dvar_a0}$b$ shows that the switching distance scales as $d_c\sim \alpha^{-1}$ which validates the scaling arguments presented above.

\subsection{Synchronization effects: comparison with nonlinear results}

\begin{figure}
\centering
\includegraphics[width=0.6\textwidth]{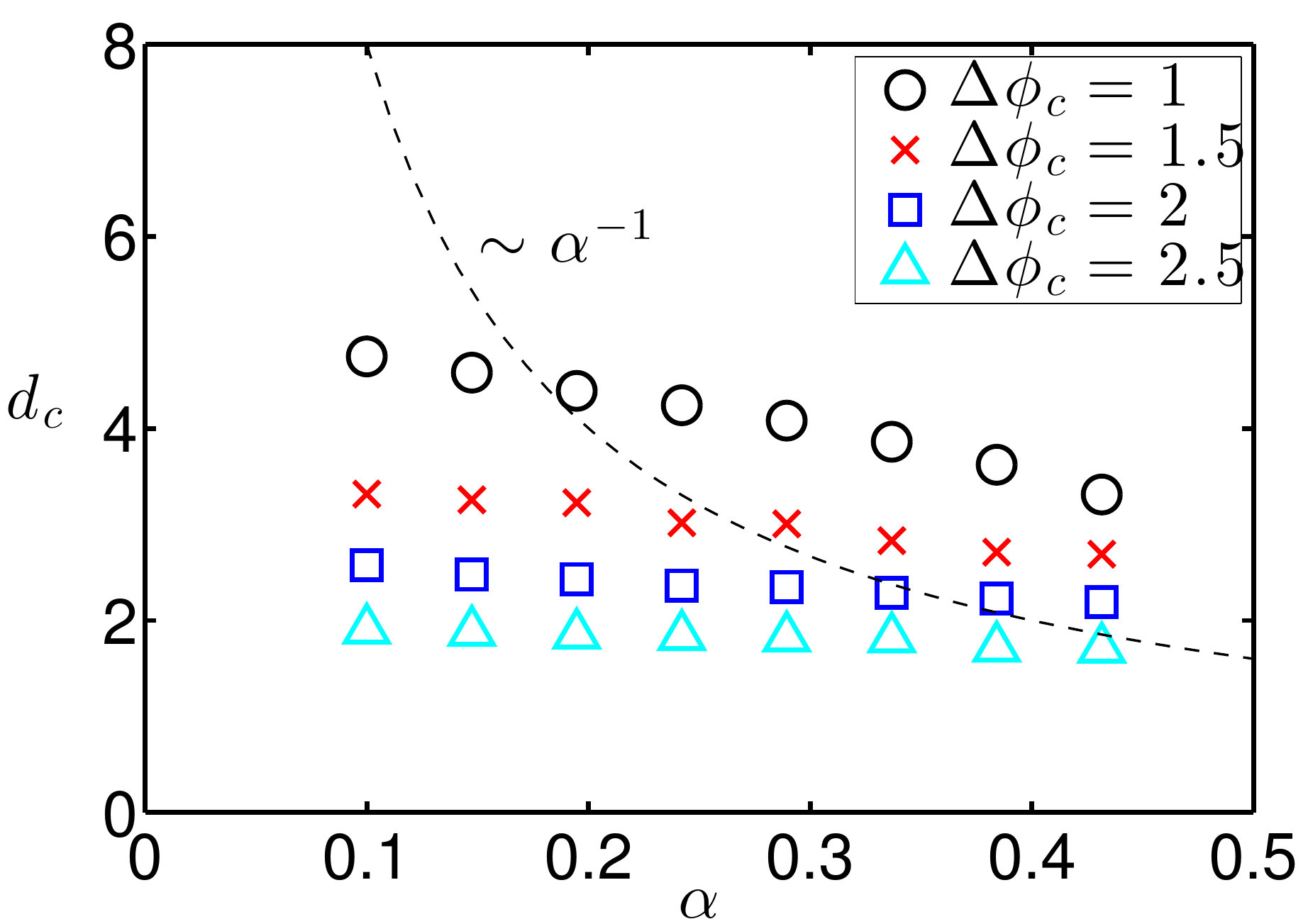}
\caption{\label{fig:dcritical_alpha_Dvar_M3_U13}Evolution with $\alpha$ of the critical distance beyond which the phase difference between the flags is lower than $\dphi_c$, for $M^*=3$, $U^*=13$. The linear result (Fig.~\ref{fig:carte_phase_M3_Uvar_Dvar_a0}) is also shown (dashed).}
\end{figure}
In the nonlinear saturated regime, $\dphi$ is not restricted to discrete values $0$ and $\pm \pi$, but instead evolves continuously with the different parameters. In order to assess the effect of $\alpha$ on the critical distance $d_c$ at which piezoelectric coupling overcomes hydrodynamics, it is necessary to define a quantitative criterion in terms of $\dphi$. Keeping all other parameters constant, the distance $d$ is varied to identify $d_c$ the distance beyond which $\dphi<\dphi_c$ for a given cut-off value $\dphi_c$. 

This distance $d_c$ is plotted as a function of $\alpha$ in Fig.~\ref{fig:dcritical_alpha_Dvar_M3_U13}, which shows that $d_c$ is indeed a decreasing function of $\alpha$. However, the evolution of $d_c$ with $\alpha$ does not follow $d_c\sim\alpha^{-1}$ in the nonlinear regime. One possible reason is that the scaling analysis proposed in Section~\ref{sec:scaling} is essentially based on a far-field analysis where the flags' relative distance is much greater than their length. This approximation is not valid in the results presented on Fig.~\ref{fig:dcritical_alpha_Dvar_M3_U13}.

\section{The role of the output circuit}\label{sec:circuit}
When electrodynamic coupling becomes dominant over hydrodynamics, it is expected that the precise design of the output circuit will play a major role in determining the synchronization of the flags~\citep{michelin:2013,xia:2015}. In the following, we investigate two effects, namely the reversal of the flags' connection and the addition of an inductance in the output loop, before concluding on the role of electrodynamic coupling by comparing our results to the case of two piezoelectric flags connected to independent circuits.

\subsection{Inverse connection (IC)}
In the inverse connection shown in Fig.~\ref{fig:circuit_R}$b$, the upper electrode of the upper flag is connected to the lower electrode of the lower one, and the dynamics of the corresponding output circuit is described by:
\begin{equation}
\label{eq:inverse_resistive}
V+\beta(\frac{\partial Q_1}{\partial t}- \frac{\partial Q_2}{\partial t})=0.
\end{equation}

\begin{figure}
\centering
\includegraphics[width=\textwidth]{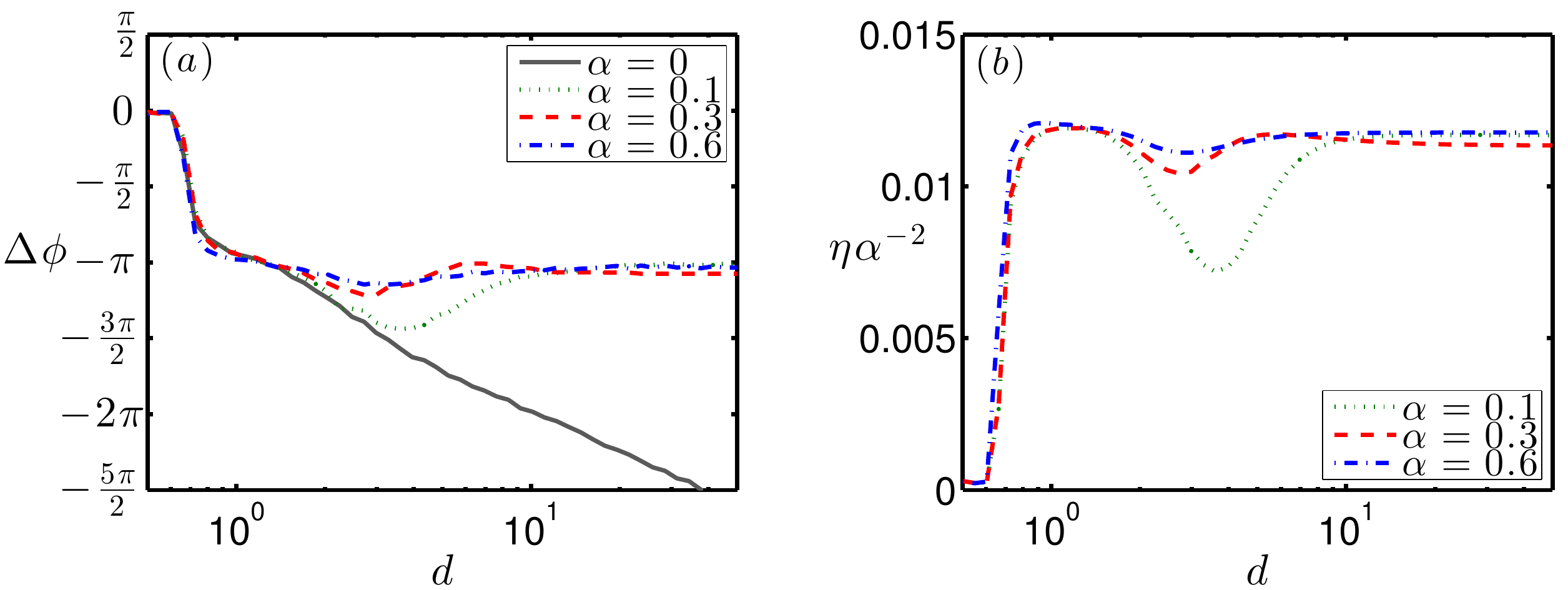}
\caption{\label{fig:inv_phase_eff_dis_Rpure_a03_a06_M3_U13_1p_para}
Evolution with the relative distance $d$ of $(a)$ the phase shift $\dphi$, and $(b)$ the rescaled harvesting efficiency $\eta\alpha^{-2}$ for $M^*=3$, $U^*=13$, $\beta=0.15$ when the flags are connected in the Inverse Connection (IC).}
\end{figure}

The voltages applied on the flag are now opposite ($V_1=-V_2=V$) and so are the piezoelectric torques: such coupling now favors an out-of-phase synchronization. This is confirmed on Figure~\ref{fig:inv_phase_eff_dis_Rpure_a03_a06_M3_U13_1p_para}: for large distances, out-of-phase synchronization is induced by the piezoelectric effect and corresponds to a maximum efficiency. In the inverse connection, the current forcing the output resistor is $\dot{Q}_1-\dot{Q}_2$: an out-of-phase flapping leads to an additive forcing of the flags on the circuit ($\dot{Q}_2=-\dot{Q}_1$).
The exact synchronization is therefore modified by the reversal of the flags' connection, but the main fundamental result still holds: at large distance, the electrodynamic coupling of the two flags overcomes hydrodynamic effects and imposes the relative phase of flapping so that it performs the most efficient transfer to the circuit.

\subsection{Resonant circuit}
\begin{figure}
\centering
\includegraphics[width=.7\textwidth]{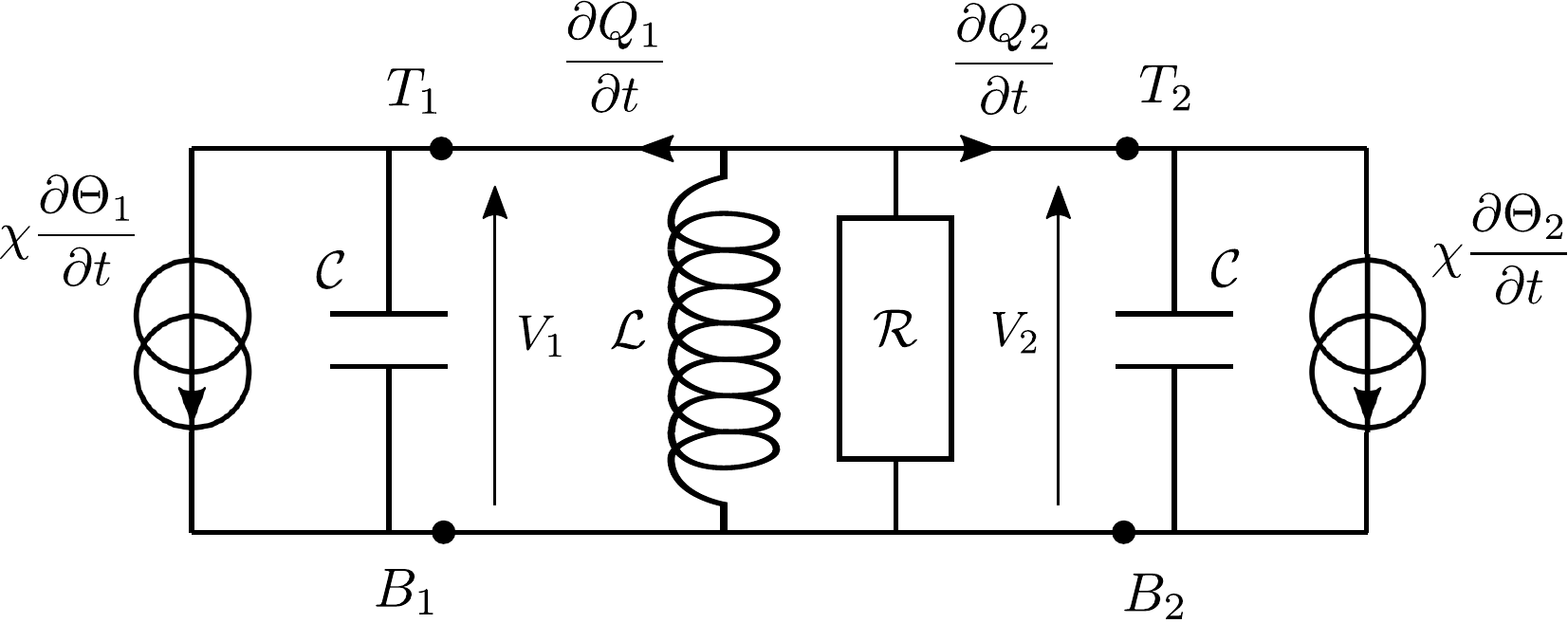}
\caption{\label{fig:circuit_RL}Resistive-inductive circuit connected to two piezoelectric flags}
\end{figure}

The addition of an inductive component to the output circuit provides the electrical system with a resonance frequency. The coupling of the flag to the resonant circuit has recently been  shown to enhance energy harvesting and can modify the flapping dynamics due to the inverse piezoelectric effect~\citep{xia:2015,xia:2015b}. In this section, the output circuit contains a resistor $\mathcal{R}$ and inductor $\mathcal{L}$ connected in parallel (Fig.~\ref{fig:circuit_RL})

\begin{figure}
\centering
\includegraphics[width=.75\textwidth]{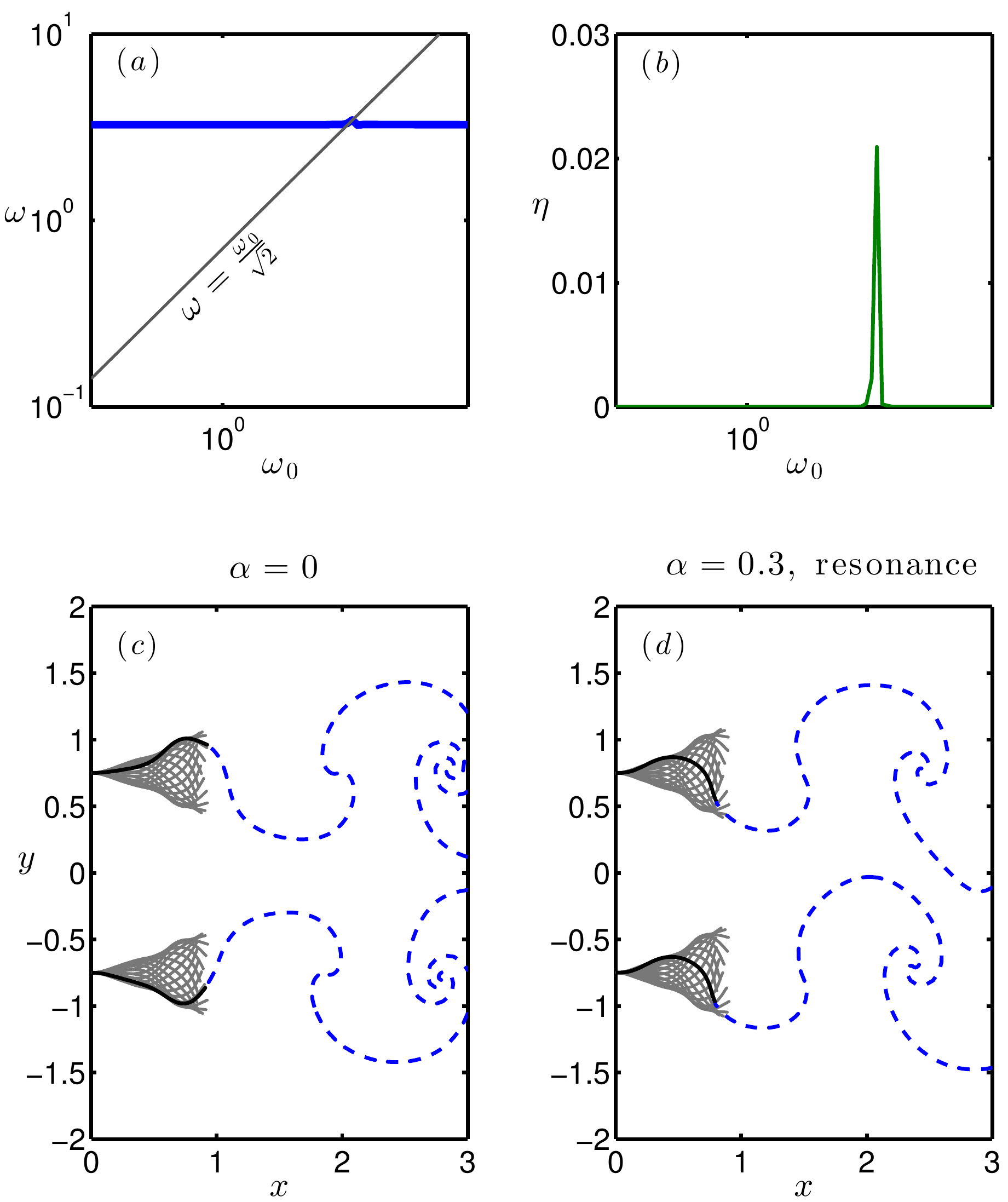}
\caption{\label{fig:eff_freq_Ovar_dis1p5_Induct_a03_M3_U13_1p_para_motions}$(a)$ Flapping frequency $\omega$ and $(b)$ harvesting efficiency $\eta$ with the normal connection, $d=1.5$, $\alpha=0.3$, $M^*=3$, $U^*=13$, $\beta=10$. Flapping motion of two flags with $d=1.5$, $M^*=3$, $U^*=13$ and $(c)$ $\alpha=0$, $(d)$ $\alpha=0.3$, $\beta=10$, $\omega_0=4.88$, normal connection at resonance.}
\end{figure}
The dimensionless equation for the circuit's dynamics can now be written as:
\begin{equation}
\label{eq:circuit_RL_ConA}
2\frac{\partial^2 V}{\partial t^2}+\frac{1}{\beta}\pd{V}{t}+\omega_0^2V+\frac{\alpha}{U^*}\left( \frac{\partial^2\Theta_1}{\partial t^2}+\pd{^2\Theta_2}{t^2}\right)=0,
\end{equation}
with $\omega_0=L/(U_\infty\sqrt{\mathcal{LC}})$.

When the flapping frequency of the flag $\omega$ matches the natural frequency of the circuit $\omega_0/\sqrt{2}$ (the circuit contains two identical capacitances), the circuit is forced at resonance leading to larger energy transfers from the mechanical system to the output circuit (Fig.~\ref{fig:eff_freq_Ovar_dis1p5_Induct_a03_M3_U13_1p_para_motions}). For the parameter values of Fig.~\ref{fig:eff_freq_Ovar_dis1p5_Induct_a03_M3_U13_1p_para_motions}, an out-of-phase flapping is observed in the absence of any piezoelectric coupling ($\alpha=0$) (see Fig.~\ref{fig:eff_freq_Ovar_dis1p5_Induct_a03_M3_U13_1p_para_motions}$c$), which would compete with the in-phase forcing introduced by the electrodynamic coupling, and should therefore result in destructive interactions and low efficiency. This is however not observed near the resonance peak, because the electrodynamic interaction of the two flags through the resonant circuit leads to a modification in their synchronization: near resonance, in-phase flapping is observed (Fig.~\ref{fig:eff_freq_Ovar_dis1p5_Induct_a03_M3_U13_1p_para_motions}$d$) with high efficiency. This can be interpreted as follows: when forced at resonance, the circuit experiences large voltage amplitude resulting in an increased feedback on the flags' dynamics through the piezoelectric torque. Forcing at resonance increases the inverse piezoelectric effect, and therefore effectively increases the electrodynamic coupling which becomes dominant over the hydrodynamic interactions. In the context of the previous section, this amounts to a reduction in the critical distance at which electrodynamic and hydrodynamic interactions balance.

\subsection{Comparison with two flags connected to different circuits}
By connecting both flags to the same circuit, the present configuration confers the circuit a double role: a coupling mechanism and an output where the energy is used. In this section, we finally attempt to shed some light on the former role by comparing our results to that obtained for two hydrodynamically-coupled flags connected to two independent and identical circuits. In that case, the harvested power can be obtained by summing the different contribution of each circuit, but the flags do not experience any electrodynamic coupling~\citep{song2014energy}. Moreover, the energy harvesting performance is not directly affected by the relative phase of flapping of the two flags, but is still impacted by the flags' interaction through the variations of their flapping amplitude or frequency depending on their hydrodynamic synchronization.

\begin{figure}
\centering
\includegraphics[width=\textwidth]{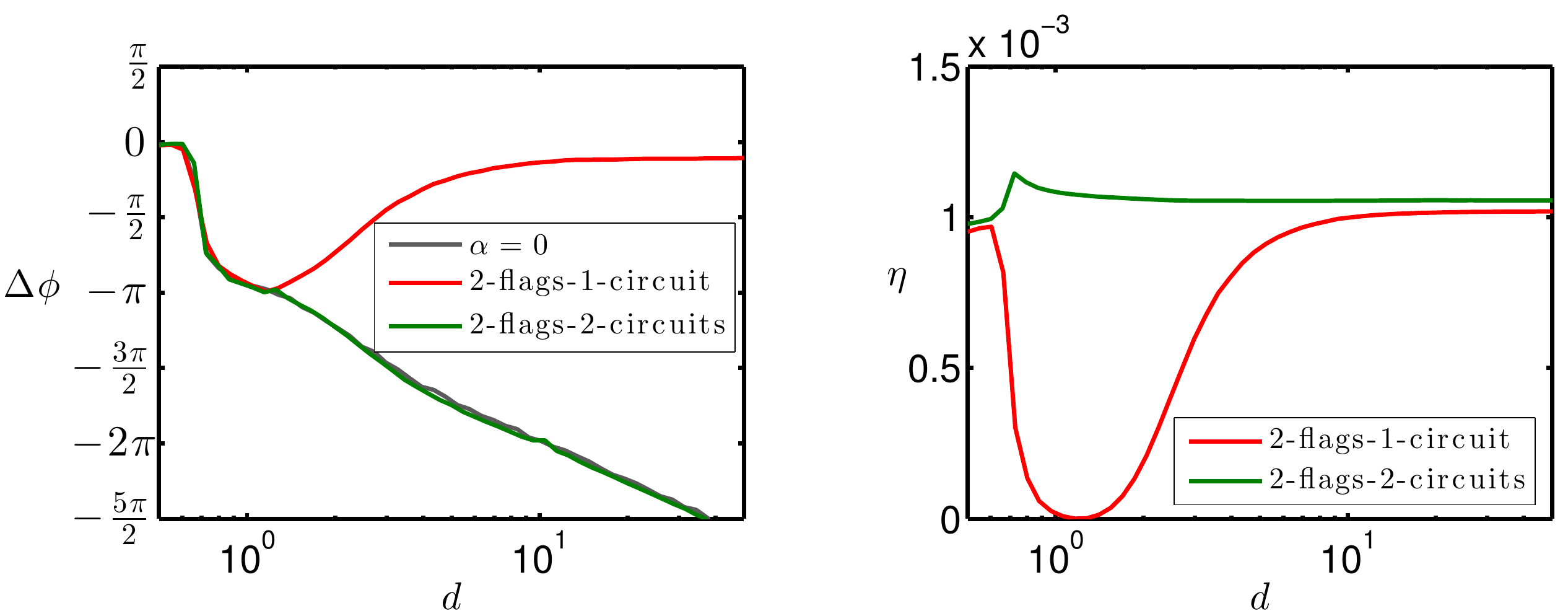}
\caption{\label{fig:eff_phase_1cir_2cir}$(a)$ Phase shift $\dphi$ and $(b)$ Harvesting efficiency $\eta$ as function of $d$ with two side-by-side flags and $M^*=3$, $U^*=13$ and $\alpha=0.3$, connected in one circuit ($\beta=0.15$) and two isolated circuits ($\beta=0.31$). In $(a)$, the phase shift obtained with $\alpha=0$ is also plotted for comparison.}
\end{figure}

Consistently, for two independent circuits, the relative phase of the flags is observed to follow exactly that of the non-piezoelectric flags ($\alpha=0$) which contrasts with the one-circuit configuration (Fig.~\ref{fig:eff_phase_1cir_2cir}$a$). Figure~\ref{fig:eff_phase_1cir_2cir}($b$) demonstrates that the electrodynamic interaction is not necessarily beneficial as it is sensitive to the flags' relative phase: for small distance, hydrodynamic drives an out-of-phase synchronization which leads to destructive interactions of the two flags in forcing the same circuit and small efficiency, while the two-circuit system maintains its efficiency regardless of the phase shift. 

The main advantage of the one-circuit configuration remains its ability to power a single larger device as it effectively synchronizes both outputs in the same electrical signal. Other techniques are however available to combine two electrical signals with different phase~\citep{chen2015energy}.

\section{Conclusion and perspectives}\label{sec:conclusions}
In this work, we investigated the competition of two different effects driving the synchronization of two side-by-side piezoelectric flags and the consequences on their energy harvesting performance: (i) the hydrodynamic coupling resulting from the fluid motion induced by each flag's displacement leading to a mechanical forcing of the other flag and (ii) the electrodynamic coupling of the flags' through the output circuit. When dominant, we have demonstrated that the latter has two main consequences: because of the inverse piezoelectric effect it can force a particular synchronization of the flags (either in-phase or out-of-phase depending on the connection used), and the resulting synchronization leads to an additive forcing of the two flags on the output circuit leading to a larger efficiency. 

Two factors have been identified to influence the relative importance of hydrodynamic and electrodynamic coupling: (i) the distance between the flags $d$ and (ii) the nature of the output circuit. The former does not impact electrodynamics but directly impacts the intensity of the hydrodynamic coupling: at larger distances, the flow perturbations introduced by one flag near the other become negligible. Hence, at larger distances hydrodynamic effects become subdominant and the flags' synchronization is driven by electrodynamic interactions. The output circuit also plays an important role: higher voltage in the output loop increases the piezoelectric feedback forcing on the structure and effectively increases electrodynamic coupling. Resonant circuits are therefore able to trigger electrodynamic synchronization of the flags at much shorter distances than resistive circuits.

We exclusively focused here on the interaction of two side-by-side flags for which hydrodynamics introduce a symmetric coupling between the flags. Other configurations deserve further scrutiny, in particular tandem flags: one flag is placed in the wake of the other, breaking the hydrodynamic coupling's scrutiny as the downstream flag experiences a much stronger forcing for its relative phase of flapping. This downstream flag also experiences a larger flapping amplitude \citep{zhang:2000, ristroph2008anomalous}, which is beneficial from an energy harvesting point of view. In that configuration, electrodynamic coupling is likely to be subdominant; it is therefore expected that much of the energy harvesting performance can be inferred from the pure hydrodynamic coupling problem. 

Our results show that a strong electrodynamic coupling is beneficial to the energy harvesting performance as it sets a particular synchronization in which the forcing of the flags on the circuit interact constructively. A possible route for improvement of the system's efficiency therefore lies within the circuit itself, for example using active circuits engineering such as Synchronized Switch Harvesting on Inductor (SSHI) to effectively synchronize the mechanical and electrical systems and enhance energy transfers~\citep{guyomard2005,pineirua2016}. 

The synchronization of more than two flags finally offer further opportunities and challenges. As demonstrated here, the relative position of the flags plays a critical role in setting their relative phase through hydrodynamic interactions. For larger flag assemblies, more complex phase distributions can be achieved~\citep{schouveiler:2009,michelin:2009}. In-depth understanding of such synchronization is however needed in order to assess the interest of such assemblies for energy harvesting purposes.

\section*{Acknowledgments}
This work was supported by the French National Research Agency ANR (Grant ANR-2012-JS09-0017). 




%\bibliographystyle{apsrev} 
%\bibliography{biblio_yifan}

%\begin{thebibliography}{00}
%%% Text of bibliographic item
%
%\bibitem{}
%
%\end{thebibliography}
\end{document}